\documentclass[11pt]{amsart}
\usepackage{geometry}                
\usepackage{graphicx}
\usepackage{amssymb}
\usepackage{epstopdf}
\title{Quantization of classical curves}
\author[X.\ Liu]{Xiaojun Liu}
\address{
Department of Applied Mathematics\\
China Agricultural University\\
Beijing, 100083, China}
\email{xjliu@cau.edu.cn}

\author[A.\ Schwarz]{Albert Schwarz}
\address{
Department of Mathematics\\
University of California\\
Davis, CA 95616, USA}
\email{schwarz@math.ucdavis.edu}

\begin{document}
\maketitle
\begin{abstract}
  We discuss the relation between quantum curves (defined as solutions of equation $[P,Q]=\hbar$,
  where $P,Q$ are ordinary differential operators) and classical curves. We illustrate this relation
  for the case of quantum curve that corresponds to the $(p,q)$-minimal model coupled to 2D gravity.

\end{abstract}
\vskip .3in

{\small{\bf Keywords:} \keywords{Quantum curve; string equation; p,q-minimal model.}

{\bf PACS No.:} 11.25pm}

\vskip .3in

It was shown by Krichever that a solution of equation $[P,Q]=0$, where $P,Q$ are ordinary
differential operators can be obtained from two meromorphic functions on an algebraic curve $X$ if
we assume that these functions obey certain conditions. He proved similar results for ordinary
differential operators with matrix coefficients and for finite difference operators.

These results led to the idea that a quantum curve can be identified with a solution of the equation
$[P,Q]=\hbar$ where $P,Q$ are ordinary differential operators \cite {M,SCH,D}.

In~Ref.~\cite{SCH} it was proven that under certain conditions the moduli space of solutions to the
equation $[P,Q]=\hbar$ can be identified with the moduli space of solutions to the equation
$[P,Q]=0$. This means that there exists a quantization procedure that is unique in some sense.  In
Ref.~\cite{KV} this result was generalized to the matrix differential operators. More detailed
consideration as well as the proof of a similar result for difference operators was given in~Ref.~\cite
{sch2}.  The proofs in these papers are based on the techniques of Sato Grassmannian.

In the present paper we will give an explanation of the quantization procedure for commuting pairs
of differential operators, describing the algorithmic part of Ref.~\cite{sch2}. We will
illustrate this procedure on some examples. We will discuss shortly the relation to Eynard-Orantin
topological recursion \cite {EO}, to Gukov-Su{\l}kowski quantization, \cite {GS} and to the theory
of quantum curves developed in Ref.~\cite {D}.

Let us consider the algebra $\mathcal D$ of ordinary differential operators $\sum a_k(x)D^k$ where
$D=\frac{d}{dx} $ and the coefficients $a_k(x)$ are formal power series with respect to $x$. This
algebra acts in natural way on the space $\mathcal H$ of Laurent series $\sum c_nz^n$; namely if
$f(z)\in\mathcal H$ then $D f(z) := z f(z)$ and $x f(z):= -\partial_zf(z)$.

This action can be extended to pseudodifferential operators (i.e. in the expression for the operator
we can allow negative $k$).

The subspace $\mathcal H_+$ consisting of polynomials is invariant with respect to the action of
differential operators; moreover, differential operators can be characterized as pseudodifferential
operators preserving this subspace.

Let us fix an operator $Q\in \mathcal{D}$. We say that the elements $u_1,...,u_q$ form a $Q$-basis
of $\mathcal H_+$ iff the elements $Q^ku_i$ form a basis of $\mathcal H_+$. For example, if $Q$ is a
monic operator of order $q$ one can take a $Q$-basis $u_i=z^{i-1}$.  The $Q$-basis is defined up to
multiplication by an invertible matrix having polynomials of $Q$ as entries. The group of such
matrices will be denoted by $G$.

If $P\in \mathcal{D}$ is another differential operator we define the companion matrix $M$ associated
to the pair $(P,Q)$ as the matrix of the coefficients in the expression
\begin{equation}
\label{m}
Pu_i=M_i^j(Q)u_j.
\end{equation}
The entries of this matrix are polynomials with respect to $Q.$ The matrix $M$ depends on the choice
of the $Q$-basis. If $[P,Q]=\hbar$  and the basis $u_i$ is replaced by the basis $g^j_i(Q)u_j$ then the matrix $M$ should be replaced by
$gMg^{-1}+\hbar \frac{dg}{dQ}g^{-1}.$ Notice that this is the standard formula for gauge transformation, hence the companion matrix cam be regarded as a connection.

Using this notion we can define the quantization of a pair $(P,Q)$ of commuting differential
operators where $Q$ is a monic differential operator. We say that the quantization leads to the pair $(P_{\hbar},Q_{\hbar})$ of differential
operators obeying $[P_{\hbar},Q_{\hbar}]=\hbar$ and having the same companion matrix in the $Q$-basis $u_i=z^{i-1}$ where $1\leq i\leq q$.

If the operator $Q$ is a normalized operator of order $q$ (i.e. the leading term is $\partial ^q$
and the subleading term vanishes) one can find such a pseudodifferendial operator $S$ of the form
$S=1+ \sum_{k<0} s_k(x)D^k$ that $S^{-1}QS=D ^q$. The natural projection of the space $V=S\mathcal
H_+$ onto $\mathcal H_+$ is an isomorphism (i.e. $V$ belongs to the big cell of infinite-dimensional
Grassmannian). Using this remark one can check that to solve the equation
$[P_{\hbar},Q_{\hbar}]=\hbar$ we should find a space $V_{\hbar}$ belonging to the big cell of
Grassmannian that is invariant with respect to the multiplication by $z^q$ and with respect to an
operator $\tilde P_{\hbar}$ having the form $\hbar\frac{d}{dz^q}+s_{\hbar}(z)$ where $s_{\hbar}(z)$
stands for the operator of multiplication by a Laurent series denoted by the same symbol and
$\frac{d}{dz^q}$ should be understood as $\frac{1}{qz^{q-1}}\frac{d}{dz}$ (see Ref.~\cite {SCH}).

The physical quantities can be expressed in terms of the tau-function corresponding to $V_{\hbar}$
or in terms of corresponding vector $\Psi_{\hbar}$ in the fermionic Fock space. Notice that one can
find the vector $\Psi_{\hbar}$ using the following statement: if the subspace $V$ obeys $AV\subset
V$ where $A$ is a linear operator in $\mathcal H$ having the matrix $a_{mn}$ in the standard basis
then the corresponding vector in the fermionic Fock space is an eigenvector of the operator $\sum
a_{mn}:\psi_m\psi_n^+:$ (see Ref.~\cite {KS}).

The  companion matrix of  the pair $(P_{\hbar},Q_{\hbar})$ can be described as the matrix
of $\tilde P_{\hbar}$ in the $z^q$-basis $v_i$ of $V_{\hbar}$:
\begin{equation}
\label{mm}
\tilde P_{\hbar}v_i= M_i^j(z^q)v_j.
\end{equation} Notice that in this approach the entries of this matrix are polynomials of $z^q.$ 

The space $V_{\hbar}$ has a natural $z^q$-basis 
$$v_i=z^i +\text{lower order terms},$$ 
here $0\leq i<q.$ This basis (and the corresponding matrix $M$) are defined up to triangular
transformation with constant coefficients. One can say that the operators  $(P_{\hbar},Q_{\hbar})$ are
obtained by means of quantization of $(P_0, Q_0)$ if they have the same companion matrix in the natural $z^q$- basis of  $V_{\hbar}.$ (This definition
agrees with the definition in terms of $Q$-basis $z^i$ of $\mathcal H_+$ because the operator $S$ transforms this basis into the natural basis of $V_{\hbar}.$ )

Sometimes it is convenient to write the equation (\ref {mm}) in the form

\begin{equation}
\label{b}
\left(\hbar\frac{d}{dz^q}+s_{\hbar}(z)\right)u_i(z)=  B_i^j(z)u_j
\end{equation}
where $v_i=z^iu_i$ and
$$B=(B_i^j(z)) = \left(M_i^j(z^q) z^{j-i}-\frac{i\hbar}{q z^q}\delta_i^j\right).$$

To find $s_{\hbar}(z)$ we use the condition that (\ref {b}) considered as an equation for $u_i$
should have solution with asymptotic behavior $u_i=1+\cdots $ where ... stands for lower order
terms. It is convenient to solve at first the following auxiliary equation
\begin{equation}
\label{a}
\hbar\frac{d}{dz^q}w_i(z)=  B_i^j(z)w_j.
\end{equation}
Notice that we can introduce the covariant derivative (meromorphic connection on $\mathbb{C}$) by
the formula
$$\nabla=\hbar\frac{d}{dz^q}-B_i^j(z),$$
then the equation (\ref {a}) specifies flat sections. Notice that that the connection (\ref {a}) is gauge equivalent to the connection specified by the matrix $M$; we
could work in terms of the latter connection.

We express $s_{\hbar}(z)$ in terms of formal diagonalization of (\ref {a}). We should solve the
following problem: Find a formal change of variables of the form $w_i(z)=R_i^j(z)t_j(z)$ that
diagonalizes the equation i.e. reduces it to the form
\begin{equation}
\label{d}
\hbar\frac{d}{dz^q}t_i(z)=  \Lambda_i (z)t_i(z).
\end{equation}
It is well known that such a diagonalization is possible if the leading term of the matrix $B$ has
$q$ distinct eigenvalues. \cite {WAS} Using this statement we can prove that there exist $q$
different solutions to (\ref{b}) corresponding to the coefficients $\Lambda_i(z)$; namely, we can
take
\begin{equation}
\label{ }
s_{\hbar}(z)=\Lambda_i(z).
\end{equation}
To prove this fact we notice first of all that after the change of variables we obtain the equation 
\begin{equation}
\label{dd}
\hbar\frac{d}{dz^q}t_i(z)=  \Lambda_i ^j(z)t_j(z),
\end{equation}
where 
$$\Lambda_i ^j(z)=
S_i^rB_r^mR^j_m-\hbar S_i^m\frac{dR_m^j}{dz^q}.$$ 
( Here $S$ denotes the matrix inverse to the matrix $R$.)

In other words,
\begin{equation}
\label{ddd}
R^i_k\Lambda_i^j(z)=B_k^mR^j_m-\hbar \frac{dR_k^j}{dz^q}.
\end{equation}

We choose $R$ in such a way that the matrix $\Lambda$ is a diagonal matrix with entries $\Lambda
_i.$ Then it follows from (\ref {ddd}) that for every $r$ the series $w_i(z)=R_i^r(z)$ satisfy the
equation (\ref {b}) with $s_{\hbar}(z)=\Lambda_r(z).$

Note that the companion matrix $M(z^q)$ is obviously invariant with respect to transformations $z\to
\epsilon z$ where $\epsilon ^q=1.$ It follows that the group $C_q$ of $q$-th roots of $1$ is a
symmetry group of the equation (\ref {b}). It acts on the coefficients of (\ref{d}): if $\Lambda(z)$
is one of these coefficients then 
$\Lambda (\epsilon z)$ is also a coefficient.

Up to terms tending to zero as $\hbar\to 0$ the coefficients $\Lambda_i(z)$ coincide with the
eigenvalues of the matrix $B$,
or with the eigenvalues $\Lambda_i^{(0)}$ of the matrix $M$ ( up to terms of order
$\hbar$ matrices $B$ and $M$ are similar ).

Let us denote by $B^{(i)}$ the $i$-th order term of matrix $B$ as an expansion in $\hbar.$ If
$R^{(0)}$ is the matrix of eigenvectors of matrix $B^{(0)}$, i.e. $(R^{(0)})^{-1}B^{(0)}
R^{(0)}:=\Lambda^{(0)}(z)$ is a diagonal matrix with diagonal elements being eigenvalues
$\Lambda_i^{(0)}$ then
\begin{equation}\label{eq:Lambda} 
  \Lambda(z) = \Lambda^{(0)}(z) +
  \left[(R^{(0)})^{-1}B^{(1)}R^{(0)}-(R^{(0)})^{-1}\frac{d}{d(z^q)}R^{(0)}\right]^{\text{diag}}\hbar+o(\hbar). 
\end{equation} 

Note that for $\hbar=0$ the commuting operators $P_0,Q_0$ satisfy the algebraic equation
$A(P_0,Q_0)=0$, where $A$ stands for the characteristic polynomial of the matrix $M$:
$$A(P_0,Q_0)=\det (P_0I -M(Q_0)).$$
For $\hbar\neq 0$ one can find an operator annihilating $v_0.$ To do this we should exclude
$v_1,\cdots, v_{q-1}$ from (\ref {mm}) or $u_1,\cdots, u_{q-1}$ from (\ref{b}). 

It is easier to find an operator  $\hat A$ annihilating $w_0=\rho(z)v_0$:
$$\hat A w_0=0,$$ where $\rho(z) =\exp(\hbar^{-1}\int s_\hbar(z) d(z^q))$. 
There is a standard way to construct $\hat A$ as a differential operator with meromorphic
coefficients (i.e. as a polynomial with respect to $\partial _z$ with coefficient that are
meromorphic with respect to $z).$ \footnote {The meromorphic connection determined by the companion
  matrix is flat, hence we can say that it specifies a $D$-module. This $D$-module corresponds to
  one differential equation with meromorphic coefficients; in the construction of $\hat A$ we use
  this fact. This remark gives an explanation of relation between our constructions and the
  constructions of Ref. \cite {D}. } Namely, we should consider $w=(w_0,\cdots,w_{q-1}) $ as an
element of $\mathcal{F}^q$ where $\mathcal{F}$ denotes the field of meromorphic functions. Then
$w_0=<e_0,w>$ where $e_0=(1,0,\cdots,0)$ and $<...> $ denotes the standard bilinear inner product
with values in $\mathcal{F}$ (i.e. $<a,b>=\sum a_ib_i$). Defining $\nabla_*$ by the formula
$$\nabla_*=\hbar\frac{d}{dz^q}+B_i^j(z),$$
and using $\nabla w=0$ we obtain that
$$(\hbar\frac{d}{dz^q})^s w_0= <\nabla_*^s e_0, w>.$$
To find $\hat A$ we notice that the vectors $\nabla_*^s e_0$ with $s=0,\cdots q$ are linearly
dependent in $q$-dimensional vector space over $\mathcal{F}$. If we can take $$\nabla_*^q
e_0=\sum_{0\leq s<q} a_s(z,\hbar)\nabla_*^s e_0$$ then
\begin{equation}
\label{ha}
\hat A=(\hbar\frac{d}{dz^q})^q- \sum_{0\leq s<q} a_s(z,\hbar)(\hbar\frac{d}{dz^q})^s.
\end{equation}

As we have noticed another way to construct $\hat A$ is to exclude $v_1,\cdots,
v_{q-1}$ 
from (\ref {mm}); if this procedure leads to differential operator with meromorphic coefficients we
obtain an equivalent result.
 
The operator  $\hat A$ can be regarded as quantization of the classical
observable $A$; compare with Ref. \cite {GS}. 

Let us illustrate the construction of the operator $\hat A$ in terms of the matrix $M$ in the case
when $q=2.$ We start with the equation
\begin{subequations}\label{2}
\begin{align} 
(\hbar\frac{d}{dz^2}+s_\hbar(z))v_0&=av_0+bv_1,\\
(\hbar\frac{d}{dz^2}+s_\hbar(z))v_1&=cv_0+dv_1,
\end{align}
\end{subequations}
where $a,b,c,d$ are the entries of matrix $M$ (they are polynomial functions of $z^2$.)  The
function $s_{\hbar}(z)$ can be expressed in terms of the eigenvalues and eigenvectors of the matrix
$$B=
\begin{pmatrix}
  a & z b \\ z^{-1}c & d
\end{pmatrix}-
\begin{pmatrix}
  0 & 0\\ 0 & \frac{1}{2 z^2}
\end{pmatrix}\hbar 
$$ 
up to terms
of order $\hbar ^2$:
$$s_{\hbar}(z)=\frac{a+d\mp\sqrt{\Delta}}{2} -
\left(
\frac{c}{\sqrt{\Delta}}
\frac{d}{d(z^2)}
\left(\frac{\sqrt{\Delta}\mp(a-d)}{2c}\right)
+\frac1{2z^2}
\right)\hbar+o(\hbar),$$
where $\Delta = (a-d)^2+4bc$.

The commuting differential operators $P_0,Q_0$ with companion matrix  $M$ obey the characteristic
equation of this matrix : $A(P_0,Q_0)=0$, where
$$A(x,y)=x^2-(a+d)x+ad-bc.$$ This means that
$$s_{\hbar}(x) =x(y)+O({\hbar})$$
where $x(y)$ stands for one of branches of the expression for $x$ obtained from the equation $A(x,y)=0.$

Excluding $v_1$ from (\ref {2}) we obtain that 
\begin{displaymath}
  \hat{A} (\rho(z) v_0) = 0,
\end{displaymath}
where \begin{equation}
\label{22}
\hat{A} = \left(\hat{x}^2 - (a+d)\hat{x} + (ad-bc)\right) - \hbar b^{-1}\left(b' \hat{x} - 
  (b'a-ba')\right),
  \end{equation}
  $\rho(z) = \exp\left(\hbar^{-1}\int s_\hbar(z) d(z^2)\right),$
$\hat{x}=\hbar\frac{d}{d(z^2)}$, $\hat{y}=z^2$, $a$,$b$,$c$,$d$ are polynomials in $\hat{y}$.

The calculation based on the formula (\ref {ha}) leads to the same expression.

We took as a starting point for construction of quantum curve a pair of commuting differential
operators $P_0,Q_0$. Such pair of operators obeys an algebraic equation, therefore they can be
considered as meromorphic functions on algebraic curve.  Let us show that one can start with a pair
of meromorphic functions $u,v$ on the algebraic curve $C$ of genus $g$ in the construction of
quantum curve. This is reminiscent of Eynard-Orantin topological recursion \cite {EO}, but our
conditions on functions $u,v$ are different.

For simplicity let us consider the case when the functions $u,v$ have only one pole located at
non-singular point $c\in C$. Let us consider a subspace $W$ of the space of meromorphic functions on
$C$, that is invariant with respect to multiplication by $u$ and $v$ and has a basis $s_n$ such that
$s_n$ has a pole of order $n$ at the point $c.$ \footnote {If we try to use as $W$ the space of
  meromorphic functions having a pole only at $c$ we discover that there are $g$ gaps (Weierstrass
  gaps) in the sequence of $n$'s serving as the order of poles. Therefore we should weaken our
  conditions allowing poles elsewhere. One can allow poles at non-special divisor of degree $g-1.$}

The multiplication by $u$ and $v$ specifies commuting operators on $W$. The natural identification
of $W$ and $\mathcal{H}_+$ allows us to consider these operators as commuting differential
operators. We can use these operators as an input in the construction of quantum curve. However,
this is not necessary: one can construct the quantum curve directly from $u$ and $v$. Namely, we
should take a $v$-basis $f_1, ...f_q$ in $W$. (This means that the functions $v^kf_i$ form a basis
in $W$. The number $q$ is equal to the order of the pole of $v$.)  Using this basis we can construct
a matrix of the operator of multiplication by $u$:
$$uf_i=M_i^j(v)f_j.$$
We use this matrix to construct a quantum curve, i.e. a pair of differential operators with the
commutator equal to $\hbar$ having the matrix as the companion matrix.

Until now we have considered scalar differential operators. However, with small modifications we can
apply our considerations to matrix differential operators. In this case we should replace the space
$\mathcal{H}$ by the space $\mathcal{H}\otimes \mathbb{C}^r$ of vector-valued Laurent
polynomials. The entries of the companion matrix associated become $r\times r$ matrices.  The
function $v$ should have $r$ poles.


\section*{Examples}

In the examples below we start with functions $u,v$ defined on the algebraic curve specified by the 
equation $A(u,v)=0.$


\subsection*{1. $u^2-v=0$.}  Functions $u$ and $v$ can be parametrized as meromorphic
functions $u=z$ and $v=z^2$ on $\mathbb{P}^1$. Then the space $W=\mathcal{H}_+$ is the space of all
meromorphic functions only have poles at $\infty$. The $v$-basis for $W$ is given by $f_0=1$ and
$f_1=z$. Thus we can construct the matrix $(M^j_i(z))$ of the operator of $u$-multiplication on the
basis $\{f_i\}$
\begin{displaymath}
  M = 
  \begin{pmatrix}
    0 & 1 \\ z^2 & 0
  \end{pmatrix}.
\end{displaymath}
Then the operator $\hat A$ can be constructed according (\ref {22}).
\begin{displaymath}
  \hat A(v_0) = \left(\hat u^2-\hat v\right)(v_0) = 0,
\end{displaymath}
where $\hat u= \hbar\frac{d}{d(z^2)}\pm z- 
\frac{\hbar}{4z^2}$, $\hat v = z^2$.

\subsection*{2. $u^3-v^2=0$.} Functions $u$ and $v$ can be parametrized as $u=z^2$ and $v=z^3$. The space $W =
\mathcal{H}_+$ is the same as Example 1. The $v$-basis for $W$ is given by $f_0 = 1$, $f_1 = z$ and
$f_2 = z^2$. Thus the matrix of $u$-multiplication on this $v$-basis would be
\begin{displaymath}
  M =
  \begin{pmatrix}
    0 & 0 & 1\\
    z^3 & 0 & 0\\
    0 & z^3 & 0
  \end{pmatrix}.
\end{displaymath}
The operator $\hat A$ can be constructed according to (\ref{ha}) by finding coefficients
$a_s(z,\hbar)$ of $\nabla_*^3(e_0)$ in terms of $\nabla_*^i(e_0)$ ($0\le i< 3$), where $\nabla_*$ is
defined as $\hbar\frac{d}{d(z^q)}+M_i^j(z)$.
\begin{displaymath}
  \hat A (v_0) = (\hat{u}^3 - \hat{v}^{-1}\hbar\hat{u}^2 -\hat{v}^2)(v_0) =0,
\end{displaymath}
where $\hat{u} = \hbar\frac{d}{d(z^3)} +s_{\hbar }(z)$, $\hat{v} = z^3$.  One can take  $s_{\hbar
}(z)=z^2 - \frac{\hbar}{3z^3}.$ Other possibilities are $s_\hbar(\epsilon z)=\epsilon^2 z^2 - \frac{\hbar}{3z^3}$ and
$s_\hbar(\epsilon^2 z) = \epsilon z^2 - \frac{\hbar}{3z^3}$ where $\epsilon = e^{2\pi i/3}$.

Another way to construct the operator $\hat A$ is to exclude $v_1,v_2$ from equations  (\ref {mm}):
\begin{displaymath}
  \tilde{P}_\hbar
  \begin{pmatrix}
    v_0\\ v_1\\v_2
  \end{pmatrix}
  =
  \begin{pmatrix}
    0 & 0 & 1\\ z^3 & 0 & 0 \\0 & z^3 & 0
  \end{pmatrix}
  \begin{pmatrix}
    v_0 \\ v_1 \\v_2
  \end{pmatrix}
  =
  \begin{pmatrix}
    v_2 \\ z^3 v_0\\ z^3 v_1
  \end{pmatrix}.
\end{displaymath}
We obtain
\begin{displaymath}
  (\hat{v}^{-1}\hat{u})(\hat{v}^{-1}\hat{u})\hat{u} v_0 = v_0,
\end{displaymath}
where $\hat{u}=\tilde{P}_\hbar$, $\hat{v} = z^3$. Two results are equivalent.  (They differ by a factor 
 $\hat{v}^2$ .)


 \subsection*{3. General case $u^q-v^p=0$ where $p$ and $q$ are coprime.}  Functions $u$ and $v$ are
 parametrized as $u=z^p$, $v=z^q$.  The space $W=\mathcal{H}_+$ is the same as Example 1 and 2. The
 $v$-basis is given by $\{z^i\}_{0\le i<q}$.  The matrix $M$ of $u$-multiplication on the $v$-basis
 is
\begin{displaymath}
  M = (M_i^j) = \left(v^{\frac{p+i-\sigma(i)}{q}}\delta_{\sigma(i)}^j\right),\quad (0\le i, j< q),
\end{displaymath}
where $\sigma$ denotes the permutation on $q$-indices $\{0,1,\ldots,q-1\}$ induced by shifting each
index by $p$, in the sense of module $q$, namely $\sigma(i)\equiv p+i \mod q$.  Then the matrix $M$
can be considered as the companion matrix of operator $\tilde{P}_\hbar$ in the $z^q$-basis $v_i$
according to equation (\ref{mm}). The differential operator $\tilde{P}_\hbar$ is written by
$\hbar\frac{d}{dz^q}+s_\hbar(z)$. To find $s_\hbar(z)$, it is convenient to use equation
(\ref{b}). In this case
\begin{displaymath}
  B = (B_i^j(z)) = \left(z^p \delta_{\sigma(i)}^j - \hbar \frac{i}{q z^q}\delta_i^j\right) :=
  A_0 z^p - \hbar A_1 z^{-q},
\end{displaymath}
where $A_0 = (\delta_{\sigma(i)}^j)$, $A_1 = (\frac{i}q \delta_i^j)$. Then $s_\hbar(z)$ is given by
(\ref{ }), where $\Lambda(z)$ is defined in (\ref{eq:Lambda}). $R^{(0)}$ is the matrix of
eigenvectors of $A_0 z^p$. Since $A_0$ is the matrix of permutation $\sigma$, it is easy to see
$R^{(0)}$ can be described by Vandermonde matrix $(x_j^i)$, where $x_j=\epsilon^j$,
$\epsilon=e^{2\pi i/q}$. And the diagonal matrix $\Lambda^{(0)}(z)=(z\epsilon^i)^p\delta_i^j$ 
is consisted of corresponding eigenvalues.  If we denote the inverse matrix of $R^{(0)}$ by
$P=(P_j^i)$, then the matrix multiplication $P R^{(0)}=I$ can be considered
as the evaluation of polynomials $P^{i}(x):=\sum_{l=0}^{q-1} P_l^i x^l$ at $x_j$, such that
$P^i(x_j)=\delta_j^i$. Therefore $P^i(x)$ can be calculated by means of Lagrange interpolation
formula:
\begin{displaymath}
  P^i(x) = \frac{\prod_{k\neq i}(x-x_k)}{\prod_{k\neq i}(x_i-x_k)}= \frac{x^q-1}{q x_i^{q-1}(x-x_i)}.
\end{displaymath}
Then the $i$-th diagonal elements of the conjugation $-P A_1 z^{-q} R^{(0)}$ can be calculated
\begin{align*}
   -\frac{1}{q z^q}\sum_{k}k P_k^i x_i^k  
  = \frac{x_i}{q z^q}\frac{d}{dx}\bigg|_{x=x_i}P^i(x) 
  =-\frac{x_i^2}{q^2 z^q}\lim_{x\to x_i}\frac{(q-1)x^q - qx^{q-1}x_i + 1}{(x-x_i)^2}
  =\frac{1-q}{2q z^q}.
\end{align*}
And in (\ref{eq:Lambda}) $\frac{d}{d(z^q)}R^{(0)}$ vanishes because $R^{(0)}$ does not dependent on
$z$. Thus we obtained $q$ values of $s_{\hbar}$ corresponding to eigenvalues of $A_0$: for the
eigenvalue $1$ we get $s_\hbar(z) = z^p + \frac{1-q}{2q z^q}\hbar + o(\hbar)$, and for eigenvalues
$\epsilon^i$ we get $s_\hbar= (\epsilon^i z)^p + \frac{1-q}{2q z^q}\hbar + o(\hbar)$, ($0<i<q$). 

Let us prove that the higher order ($o(\hbar)$) terms vanish. One of possible ways to give a proof
is to check that the equations (2) or (3) have solutions that satisfy the necessary
conditions.\footnote {We should prove that there exists a desired solution for every eigenvalue of
  matrix $A_0$, however, it is sufficient to consider one of this eigenvalues; we will give the
  proof for the eigenvalue $1.$ (The group $C_q$ of $q$-th roots of unity is a symmetry group of the
  equation (\ref {b}); it acts transitively on the eigenvalues of $A_0$.)}


We will prove that for $s_\hbar(z) = z^p + \frac{1-q}{2q z^q}\hbar$ there exists a solution
$u=(u_i)$ for (\ref{b}) having the desired asymptotic behavior: $u_i = 1+ \cdots$, where $\cdots$
means lower order terms in $z$.

Let's denote $\alpha=\frac{1-q}{2q}$, so that $s_\hbar(z) = z^p + \hbar \alpha z^{-q}$. If we
express $u$ as a Laurent series $u=(u_i)=\sum_{j\ge0} \xi_j z^{-j}$ with $q$-dimensional vectors
$\xi_i$ as coefficients, then according to (\ref{b}) we have $\xi_0=(1,\cdots,1)^T$ and $\xi_j$
should satisfy the following formula:
\begin{align}
  &(A_0 - 1) \xi_i = 0 & (0\le i < p+q),\label{eq:recur1}\\
  &(A_0-1)\xi_i = \hbar\left(A_1+\alpha - \frac{i-p-q}{q}\right)\xi_{i-p-q} & (i\ge p+q).\label{eq:recur2}
\end{align}
We can show that (\ref{eq:recur1}) and (\ref{eq:recur2}) determine $\xi_i$ completely and uniquely.

As we noticed  $1$ is  one of  $q$ different eigenvalues of $A_0$. Let's denote its
one-dimensional eigenspace by $V_1$ (it is spanned by $\xi_0$). Then for $i=1,\cdots, p+q-1$, we
have $\xi_i\in V_1$. They can be determined later by using (\ref{eq:recur2}).

It is also useful to note that image space of linear transformation $A_0-1$ has codimension $1$,
its standard orthogonal complement space is $V_1$. Therefore we have an orthogonal decomposition
$\mathbb{R}^n = \mathrm{Im}(A_0-1)\oplus V_1$.

Then for $i=p+q$, we can verify that $\hbar(A_1+\alpha)\xi_0\in \mathrm{Im}(A_0-1)$, because it is
orthogonal to $\xi_0$ because of the value of $\alpha$.\footnote{Note that the value of $\alpha$ is
  crucial here. In this value changes, then $\hbar(A_1+\alpha)\xi_0\not\in \mathrm{Im}(A_0-1)$, the
  recursion formula would not hold.} 
 Then (\ref{eq:recur2}) makes sense and
$\xi_{p+q}$ can be solved up to an element in $V_1$. Let's write $\xi_{p+q}=\tilde{\xi}_{p+q} +
c_{p+q}\xi_0$, according to the decomposition $\mathbb{R}^n = \mathrm{Im}(A_0-1)\oplus V_1$. Then
$\tilde{\xi}_{p+q}$ is completely determined. The scalar $c_{p+q}$ can be determined later.

For $i = p+q+s$, ($0< s<p+q$), (\ref{eq:recur2}) can be written as
\begin{displaymath}
  (A_0-1)\xi_{p+q+s} -\hbar\left(A_1+\alpha\right)\xi_s = -\hbar\frac{s}{q}\xi_s.
\end{displaymath}
Since $\xi_s\in V_1$, we can check the l.h.s. is in $\mathrm{Im}(A_0-1)$ too.\footnote{Again the
  value of $\alpha$ is important for the same reason.} 
But the r.h.s. is in $V_1$. Therefore both sides must be $0$, which implies $\xi_s=0$. Then
$\xi_{p+q+s}\in V_1$.

For $i= 2(p+q)$, one can use the decompositon $\xi_{p+q} = \tilde{\xi}_{p+q} + c_{p+q}\xi_0$, take
the inner product $\xi_0\cdot$ to (\ref{eq:recur2}) and get
\begin{displaymath}
  0 = \hbar\xi_0^T(A_1+\alpha-\frac{p+q}{q})(\tilde{\xi}_{p+q}+c_{p+q}\xi_0).
\end{displaymath}
Then $c_{p+q} = \xi_0^T(\frac{A_1+\alpha}{p+q}-\frac{1}q)\tilde{\xi}_{p+q}$ is determined. And it
shows that the r.h.s. of (\ref{eq:recur2}) is in $\mathrm{Im}(A_0-1)$, therefore $\xi_{2(p+q)}$ can
be determined up to an element in $V_1$, namely $\xi_{2(p+q)} = \tilde{\xi}_{2(p+q)} +
c_{2(p+q)}\xi_0$, where the first component is determined and $c_{2(p+q)}$ can be determined by
looking at the case when $i=3(p+q)$.

By using this procedure and induction, we can prove that $\xi_i = 0$ if $i\neq k(p+q)$, and
$\xi_{k(p+q)}$ can be uniquely determined. Hence the solution $u=\sum_j \xi_j z^{-j}$ is determined
uniquely.

The operator $\hat A$ can be derived from (\ref {ha}) or from (\ref{mm}).  Let's use (\ref {mm}). 

Let $\hat u$ be the operator $\tilde P_\hbar$, and $\hat v$ be the operator
of $z^q$-multiplication. Then equation (\ref{mm}) can be formulated as
\begin{displaymath}
  \hat u (v_i) = M_i^j(\hat v)( v_j ) 
  =\hat{v}^{\frac{p+i-\sigma(i)}{q}}(v_{\sigma(i)}).
\end{displaymath}
Therefore
$$\hat{v}^{-\frac{p+i-\sigma(i)}{q}} \hat u (v_i ) = v_{\sigma(i)}.$$ Then the action of
the operator $\hat{v}^{-\frac{p+i-\sigma(i)}{q}}\hat u$ is simply the permutation of
indices on $v_i$. Hence by repeating the action $q$-times, the permutation circles back to identity,
then we get
\begin{displaymath}
  \left(\hat{v}^{-\frac{p+\sigma^{q-1}(0)-\sigma^q(0)}{q}}\hat{u}\right)
  \left(\hat{v}^{-\frac{p+\sigma^{q-2}(0)-\sigma^{q-1}(0)}{q}}\hat{u}\right)
  \cdots
  \left(\hat{v}^{-\frac{p+0-\sigma(0)}{q}}\hat{u}\right)(v_0) = v_0.
\end{displaymath}
To avoid the use of negative power of $\hat{v}$, we multiply $\hat{v}^{p+q-1}$ on both sides. Then
operator  $\hat A$ is obtained:
\begin{displaymath}
  \hat A (v_0) = \left[\hat{v}^{p+q-1}(\hat{v}^{-m_q}\hat{u})\cdots (\hat{v}^{-m_1}\hat{u})-\hat{v}^{p+q-1}\right](v_0) = 0
\end{displaymath}
where $m_i = \frac{p+\sigma^{i-1}(0)-\sigma^{i}(0)}{q}$.

We have described the quantum curve obtained by quantization of the classical curve $u^q-v^p=0.$ We
have proven that the subspace $V_{\hbar}$ (the point of of Grassmannian corresponding to this
quantum curve ) is invariant with respect to the operator
$$  \tilde P_\hbar=\hbar\frac{d}{dz^q}+z^p + \frac{1-q}{2q z^q}\hbar.$$  
This means that the space $V_{\hbar}$ coincides with the point of Grassmannian correspondiong to the
$(p,q)$-minimal model coupled to 2D gravity. ( See Ref.~\cite {KS}, \cite{pq} and \cite{sch1}.)

\section*{Acknowledgments}
We would like to thank M. Mulase and M. Luu for useful discussions. X.L.  was supported by China
Scholarship Council and NSF of China No. 11201477, 11171175. A. Sch. was supported by NSF grant
DMS-0805989.

\end{document}